\documentclass[12pt,reqno]{article}
\usepackage{amsfonts}
\usepackage{amsmath}
\usepackage{amsbsy}
\usepackage{graphicx}
\usepackage{amssymb,latexsym}
\numberwithin{equation}{section}

\begin{document}

\title{Dirac equation in Kerr-Taub NUT spacetime}
\begin{titlepage}
\author{ Hakan Cebeci\footnote{E.mail:
hcebeci@anadolu.edu.tr} \\
{\small Department of Physics, Anadolu University, 26470 Eski\c{s}ehir, Turkey} \\ \\
N\"{u}lifer \"{O}zdemir\footnote{E.mail:
nozdemir@anadolu.edu.tr}\\{\small Department of Mathematics,
Anadolu University, 26470 Eski\c{s}ehir, Turkey} }

\date{ }

\maketitle

\bigskip

\begin{abstract}
\noindent We study Dirac equation in Kerr-Taub-NUT spacetime. We use Boyer-Lindquist coordinates and separate the resulting equations into radial and angular parts. We get some exact analytical solutions of the angular equations for some special cases. We also obtain the radial wave equations with an effective potential. Finally we discuss the potentials by plotting them as a function of radial distance in a physically acceptable region.
\vspace{1cm}

\end{abstract}

\end{titlepage}

\section{Introduction}

The Kerr-Taub-NUT (Newman-Unti-Tamburino) spacetime is obtained by introducing an extra non-trivial magnetic mass parameter called the "gravitomagnetic monopole moment" in the Kerr metric. It describes the spacetime of a localized stationary and axially symmetric object \cite{demianski}. The solution contains three physical parameters: The gravitational mass, which is also called gravitoelectric charge; the gravitomagnetic mass that is also identified as the NUT charge; the rotation parameter that is the angular speed per unit mass. The NUT charge produces an asymptotically non-flat spacetime in contrast to Kerr geometry that is asymptotically flat \cite{misner}. Although the Kerr-Taub-NUT spacetime has no curvature singularities, there exist conical singularities on the axis of symmetry \cite{miller}. One can get rid off conical singularities by taking a periodicity condition over the time coordinate. But, this leads to the emergence of closed time-like curves in the spacetime. It means that, in contrast to Kerr solution interpreted as a regular rotating black hole, the Kerr-Taub-NUT solution cannot be identified as a regular black hole solution due to its singularity structure. An alternative physical interpretation of Kerr-Taub-NUT spacetime can be found in \cite{bonnor} where the NUT metric is interpreted as a semi-infinite massless source of angular momentum. Despite  the fact that  Kerr-Taub-NUT solution has some unpleasing properties, it is vastly studied for exploring various physical phenomena in general relativity due to its asymptotically non-flat spacetime structure \cite{jantzen, aliev, liu, bell, gamal}.

In the present paper, we study the Dirac equation in Kerr-Taub-NUT spacetime. Dirac equation has been extensively examined in various gravitational spacetimes including Schwarzschild geometry \cite{mukhopadhyay}, Kerr-spacetime \cite{chandrasekhar1, chandrasekhar2, ranganathan, dolan}, Taub-NUT geometry \cite{comtet}, Kerr-Newman AdS black hole background \cite{belgiorno}, 4-dimensional constant-curvature black hole spacetime \cite{ghosh}, rotating Bertotti-Robinson geometry \cite{badawi}, 4-dimensional Nutku helicoid spacetime \cite{hortacsu} and open universe geometry \cite{villalba}. In some of these background spacetimes, some exact analytical solutions of massive and massless Dirac equation have been presented \cite{comtet, ghosh, badawi, hortacsu, villalba}. In \cite{belgiorno}, spectral properties of Dirac Hamiltonian are given. However in the background of rotating Kerr spacetime, exact solutions of Dirac equation have been obtained only for some special values of the parameters \cite{ranganathan, dolan}. In \cite{ranganathan}, the series solution of angular Dirac equation have been given while in \cite{dolan}, angular solutions have been presented by using spectral decomposition method in which the angular wave functions are expanded in terms of spheroidal harmonics. By this method, a three-term recursion relation is achieved and eigenvalues of the angular equations are solved.

On the other hand, in almost all these works, the separability of the Dirac equation has also been discussed. Separability was first discovered in Hamilton-Jacobi and relativistic wave equations through the pioneering works of Carter \cite{carter1, carter2}. The separability has been shown to be closely related to the existence of second order St\"{a}ckel-Killing tensors \cite{carter2}. Later on, the separability of Hamilton-Jacobi and relativistic field equations has been extended to higher dimensional spacetimes in which the St\"{a}ckel-Killing tensors are also given explicitly (see \cite{chong,frolov,qing} and the references therein). The separability of the Dirac equation however was first noticed by Chandrasekhar \cite{chandrasekhar1, chandrasekhar2}. Later, it was shown that separability of Dirac equation has also been connected with the existence of a second order Killing-Yano tensor \cite{carter3}. In close connection with these tensors, Dirac equation has been proved to be separable in general vacuum type-D spacetimes \cite{kamran}. In addition, the separability of Dirac equation has been investigated in spherically symmetric spacetimes \cite{tucker}. In \cite{page}, the separability of Dirac equation in Kerr-Newman geometry has been explicitly shown by using Boyer-Lindquist coordinates. In a recent work, authors has demonstrated the separability of massive Dirac equation in AdS-Kerr-Taub-NUT spacetimes \cite{jian}.

In this work, we obtain the set of equations by employing an axially symmetric ansatz for the Dirac spinor. Equations obtained are separated into radial and angular parts with appropriate substitutions of spinor fields. We try to solve angular equations exactly. But unfortunately, we are unable to get exact analytical solutions to general angular equations for all physical parameters. Under some restrictions implemented on the separation constant, we present some exact solutions of the equations with and without gravitomagnetic mass and rotation parameters. Indeed, they can be solved exactly in terms of hypergeometric functions for the cases where the mass of the Dirac particle is  equal to or twice the frequency of the spinor wave function. In the final part, radial equations are discussed. With some transformations on the dependent and independent field variables, wave equations with an effective potential barrier are obtained. To understand the physical behavior of the potentials, they are plotted with changing frequency and gravitomagnetic mass parameter in the physically acceptable regions.

Organization of the paper is as follows: In section 2, we present the general form of Dirac equation in exterior forms. In section 3, we obtain Dirac equation in Kerr-Taub-NUT spacetime.
In the subsections, we discuss the separability of the equations, obtain the angular and radial equations. Next, we find some exact analytical solutions of the angular equations. Finally, we study the radial wave equations and examine the behavior of the potential barriers that come out in the transformed radial equations. We end up with some comments and conclusions.

\section{Dirac equation in 4-dimensional spacetime}
We consider a $4$-dimensional spacetime manifold $M$ equipped with
a Lorentzian metric $g$ with signature $(-,+,+,+)$ and a metric
compatible connection $\nabla$. We assume that our spacetime
manifold has a spin structure group $Spin_{+} (3,1) $. It is known
that the fundamental group of Lorentzian group $SO_{+} (3,1)$ is
$\mathbb{Z}_{2}$ so that it has a universal covering group of
$Spin_{+}(3,1)$ that is the multiplicative subgroup of complex
Clifford algebra $\mathbb{C}\,l_{3,1} $.

In exterior forms, the Dirac equation can be written as \cite{dereli1}
\begin{equation}
\ast \gamma \wedge D \psi + \mu  \psi \ast 1 =0, \label{1}
\end{equation}
where $\gamma $ is $\mathbb{C}\,l_{3,1}$-valued 1-form $\gamma=
\gamma^{a} e_a$. We choose the units such that $c=1$ and $\hbar=1$.
Here $\ast$ denotes
Hodge-star operator and $\mu$ is the mass of the particle. $\{ e_a
\}$'s are the orthonormal co-frame 1-forms such that the metric $
g=\eta_{ab} e^{a}\otimes e^{b}$. $\psi $ represents
$\mathbb{C}^{4}$-valued Dirac spinor whose covariant exterior
derivative can be written as
\begin{equation}
D \psi=d\psi + \frac{1}{2} \sigma^{ab} \omega_{ab} \psi, \label{2}
\end{equation}
where $\sigma^{ab}=\frac{1}{4} [\gamma^{a},\gamma^{b}] $ and $\{\gamma^{a}\}$'s satisfy the relations
$$
\left\{\gamma^{a},\gamma^{b}
\right\}=\left(\gamma^{a}\gamma^{b}+\gamma^{b}\gamma^{a} \right) =
2 \eta^{ab}I_{4 \times 4}.$$ $\omega_{ab}$ are the connection
1-forms that satisfy Cartan structure equations
$$de^{a}+\omega^{a}\,_{b} \wedge e^{b}=T^{a}$$ where $T^{a} $
denotes torsion 2-form and metric compatibility implies that
$\omega_{ab}=-\omega_{ba}$.

Since $\mathbb{C}\,l_{3,1}$ is isomorphic to ${\cal{M}}_{4}
(\mathbb{C})$ that is the set of $4 \times 4 $ complex matrices,
we can choose the representation
$$
\gamma^{0}=i \left(
\begin{array}{cc}
0& I \\
I& 0
\end{array}
\right), \ \gamma^{1}=i \left(
\begin{array}{cc}
0& \sigma^{1} \\
- \sigma^{1}& 0
\end{array}
\right),
$$

$$
\gamma^{2}=i \left(
\begin{array}{cc}
0& \sigma^{2}\\
-\sigma^{2}& 0
\end{array}
\right), \ \gamma^{3}=i \left(
\begin{array}{cc}
0& \sigma^{3} \\
- \sigma^{3}& 0
\end{array}
\right),
$$
where $ \sigma^{i}$ are the Pauli spin matrices and $ I$ is the $ 2 \times 2$ identity matrix.

\section{Dirac Equation in Kerr-Taub-NUT Spacetime}

In this section, we examine the Dirac equation in Kerr-Taub-Nut
spacetime. In Boyer-Lindquist coordinates, Kerr-Taub-NUT spacetime
can be described by the metric with asymptotically non-flat
structure,
\begin{equation}
g = - \frac{\Delta}{\Sigma} (dt- \chi d \varphi )^{2} + \Sigma
\left(\frac{d r^{2} }{\Delta} + d \theta ^{2} \right) + \frac{\sin
^{2} \theta }{\Sigma } \left( a d t - ( r^{2} + \ell^{2} + a^{2} )
d \varphi \right)^{2} \label{3}
\end{equation}
where
$$
\Sigma = r^{2} + (\ell + a \cos \theta)^{2} , \qquad \qquad \Delta
= r^{2} - 2 M r + a^{2} - \ell^{2}
$$
and
$$
\chi = a \sin^{2} \theta - 2 \ell \cos \theta .
$$
Here, $M$ is a parameter related to physical mass of the
gravitational source. $a$ is associated with its angular momentum
per unit mass and $\ell$ denotes gravitomagnetic monopole moment
of the source. For the metric (\ref{3}), we choose the co-frame
1-forms
\begin{eqnarray}
e^{0} &=& \left( \frac{ \Delta }{\Sigma} \right)^{1/2} ( d t -
\chi d \varphi ) , \qquad \qquad e^{1} = \left( \frac { \Sigma }{
\Delta } \right)^{1/2} d r , \\ \nonumber e^{2} &=& \Sigma^{1/2} d
\theta  , \qquad \qquad e^{3} = \frac{ \sin \theta }{ \Sigma^{1/2}
} \left( a dt - ( r^{2} + \ell^{2} + a^{2} ) d \varphi \right) .
\label{4}
\end{eqnarray}
We consider that the spacetime is Levi-Civita (torsion-free) such
that connection 1-forms $\omega^{a}\,_{b}$ can be determined from
the equation
$$
d e^{a} + \omega^{a}\,_{b} \wedge e^{b} = 0
$$
which has a unique solution
\begin{equation}
\omega^{a}\,_{b} = \frac{1}{2} \left( e^{c} ( \iota^{a}\,\iota_{b}
d e_{c} ) + \iota_{b} d e^{a} - \iota^{a} d e_{b} \right) .
\label{5}
\end{equation}

Here $\iota_{a} = \iota_{X_{a}} $ are inner-product operators that
satisfy $ \iota_{b}\, e^{a} = \delta_{b}^{a} $. From equation
(\ref{5}), we can determine connection 1-forms $\omega^{a}\,_{b}$:
\begin{equation}
\begin{array}{ll}
\omega^{0}\,_{1}  =  A_2 dt + A_3 d \varphi ,  \ &
\omega^{0}\,_{2} = B_3 d \varphi, \\
\omega^{0}\,_{3} =  C_1 dr +C_2 d \theta , \ & \omega^{1}\,_{2}
= E_1 dr +E_2 d\theta , \\
\omega^{1}\,_{3} = G_3 d \varphi , \ & \omega^{2}\,_{3}  = K_2 dt
+K_3 d \varphi ,
\end{array}
\label{6}
\end{equation}
where
\begin{eqnarray}
A_2= \frac{Mr^2+2 \ell^2 r+ 2 \ell ar \cos \theta- M(\ell+ a \cos
\theta )^2}{\Sigma^2} \nonumber
\end{eqnarray}
\begin{eqnarray}
A_3 = \frac{\chi\left\{ (m-r) \Sigma -2\ell^2 r- 2 Mr^2
\right\}-2 \ell r \cos \theta (r^{2} + \ell^2+ a^2)}{\Sigma^2} \nonumber
\end{eqnarray}
\begin{eqnarray}
B_3=-\frac{\Delta^{1/2} \sin \theta (\ell+ a \cos
\theta)}{\Sigma}, \  C_1=\frac{ar \sin \theta}{\Sigma
\Delta^{1/2}}, \ C_2=-\frac{\Delta^{1/2}}{\Sigma}(\ell+ a \cos
\theta),\nonumber
\end{eqnarray}
\begin{eqnarray}
E_1=-\frac{a \sin \theta ( \ell+ a \cos \theta)}{\Delta^{1/2}
\Sigma}, \  E_2= -\frac{r\Delta^{1/2}}{\Sigma} , \  G_3= \frac{r
\sin \theta \Delta^{1/2}}{\Sigma} ,\label{7}
\end{eqnarray}
\begin{eqnarray}
K_2=\frac{\ell r^2- (\ell+ a \cos \theta)(2Mr+ \ell^2 +a \ell \cos
\theta)}{\Sigma^2} , \nonumber
\end{eqnarray}
\begin{eqnarray}
K_3= \frac{\cos \theta (r^2+ \ell^2+ a^2)(\Sigma +2\ell (\ell+ a
\cos \theta)) +\chi (2Mr+2\ell^2 )(\ell+ a \cos
\theta)}{\Sigma^2}.\nonumber
\end{eqnarray}

Since the space-time is axially symmetric, we can take
\begin{equation}
\psi=e^{-i \omega t} e^{im \varphi} \left(
\begin{array}{c}
 \psi_1
(r, \theta)\\
 \psi_2
(r, \theta)\\
 \psi_3
(r, \theta)\\
 \psi_4
(r, \theta)\\
\end{array}
\right), \label{8}
\end{equation}
where $m$ denotes azimuthal quantum number. Then we substitute (\ref{6}), (\ref{7}) and (\ref{8}) into Dirac
equation (\ref{1}) and obtain the following equations:
\begin{equation}
\left(\frac{\omega \alpha_1 +m \alpha_2}{\Delta^{1/2} \sin \theta}
\right)\psi_3 -\frac{i\Delta^{1/2}}{\Sigma ^{1/2}}\,
\frac{\partial \psi_4}{\partial r} - \frac{1}{\Sigma ^{1/2}}\,
\frac{\partial \psi_4}{\partial \theta} + \frac{i}{2} (\delta_2
+\delta_4)\psi_4 + \frac{1}{2} (\delta_1 - \delta_3) \psi_4 +\mu
\psi_1=0, \label{9}
\end{equation}
\begin{equation}
\left(\frac{-\omega \alpha_3 +m \alpha_4}{\Delta^{1/2} \sin
\theta} \right)\psi_4 -\frac{i\Delta^{1/2}}{\Sigma ^{1/2}}\,
\frac{\partial \psi_3}{\partial r} + \frac{1}{\Sigma ^{1/2}}\,
\frac{\partial \psi_3}{\partial \theta} + \frac{i}{2} (\delta_4
-\delta_2)\psi_3 + \frac{1}{2} (\delta_1+ \delta_3) \psi_3 +\mu
\psi_2=0,\label{10}
\end{equation}
\begin{equation}
\left(\frac{-\omega \alpha_3 +m \alpha_4}{\Delta^{1/2} \sin
\theta} \right)\psi_1 + \frac{i\Delta^{1/2}}{\Sigma ^{1/2}}\,
\frac{\partial \psi_2}{\partial r} + \frac{1}{\Sigma ^{1/2}}\,
\frac{\partial \psi_2}{\partial \theta} + \frac{i}{2} (\delta_2
-\delta_4)\psi_2 + \frac{1}{2} (\delta_1+ \delta_3) \psi_2 +\mu
\psi_3=0,\label{11}
\end{equation}
\begin{equation}
\left(\frac{\omega \alpha_1 +m \alpha_2}{\Delta^{1/2} \sin \theta}
\right)\psi_2 +\frac{i\Delta^{1/2}}{\Sigma ^{1/2}}\,
\frac{\partial \psi_1}{\partial r} - \frac{1}{\Sigma ^{1/2}}\,
\frac{\partial \psi_1}{\partial \theta} - \frac{i}{2} (\delta_2
+\delta_4)\psi_1 + \frac{1}{2} (\delta_1 - \delta_3) \psi_1 +\mu
\psi_4=0,\label{12}
\end{equation}
where
$$
\alpha_1=\frac{\chi \Delta^{1/2}-(r^2+a^2+\ell^2)\sin \theta
}{\Sigma^{1/2}}, \qquad \alpha_2=\frac{a \sin \theta-
\Delta^{1/2}}{\Sigma^{1/2}},
$$
$$
\alpha_3=\frac{\chi \Delta^{1/2}+(r^2+a^2+\ell^2)\sin \theta
}{\Sigma^{1/2}}, \qquad \alpha_4=\frac{a \sin \theta+
\Delta^{1/2}}{\Sigma^{1/2}},
$$
and
$$
\delta_1=- \frac{\left(l+a \cos \theta \right)}{\Sigma^{3/2}}\Delta^{1/2},
\qquad \delta_2=\frac{ar \sin \theta}{\Sigma^{3/2}},
$$
$$
\delta_3=\frac{\cos \theta \Sigma - a \sin^2 \theta (l+a \cos
\theta)}{\sin \theta \Sigma^{3/2}}, \qquad \delta_4=- \frac{\left(
\Sigma(r-M)+r \Delta \right)}{\Delta^{1/2}\Sigma^{3/2}} .
$$
\subsection{Separability of the equations}
Although in \cite{kamran}, Dirac equation is proven to be separable in the Carter class of type-D vacuum spacetimes where our Kerr-Taub NUT background also belongs to and the existence of a second order Killing-Yano tensor implies the separability of the Dirac equation, for completeness of the work and also for instructional purposes, it would be useful to discuss and explicitly illustrate the separability work by employing Boyer-Lindquist coordinates. For that purpose, we add and subtract equations (\ref{9}), (\ref{10}), (\ref{11}) and (\ref{12}) and define
\begin{eqnarray}
F_1&= & i\left( r-i (\ell+a \cos \theta)\right)^{1/2} (\psi_1 +\psi_2),\nonumber\\
F_2& =& -i\left( r-i (\ell+a \cos \theta)\right)^{1/2} (\psi_2 -\psi_1),\\
F_3& =& \left( r+i (\ell+a \cos \theta)\right)^{1/2}(\psi_3 +\psi_4),\nonumber\\
F_4& =& \left( r+i (\ell+a \cos \theta)\right)^{1/2} (\psi_4 -\psi_3)\nonumber. \label{13}
\end{eqnarray}
We then simplify the resulting equations and finally get
\begin{eqnarray}
\left\{\frac{ma- \omega(r^2+ \ell^2 +a^2)}{\Delta^{1/2}}- \cal{D}
\right\}F_3+\left\{ \frac{m-\omega \chi}{\sin \theta}
- \cal{L} \right\}F_4 \nonumber \\
- \mu \left(ir -(\ell+ a \cos \theta) \right)F_1=0,
\label{14}
\end{eqnarray}
\begin{eqnarray}
\left\{\frac{-m +\omega \chi}{\sin \theta}- \cal{L}
\right\}F_3+\left\{ \frac{-ma+\omega(r^2+ \ell^2
+a^2)}{\Delta^{1/2}}
- \cal{D} \right\}F_4 \nonumber \\
- \mu \left(ir -(\ell+ a \cos \theta) \right)F_2=0,
\label{15}
\end{eqnarray}
\begin{eqnarray}
\left\{\frac{ma- \omega(r^2+ \ell^2 +a^2)}{\Delta^{1/2}}+ \cal{D}
\right\}F_1+\left\{ \frac{m-\omega \chi}{\sin \theta}
- \cal{L} \right\}F_2 \nonumber \\
+ \mu \left(ir +(\ell+ a \cos \theta) \right)F_3=0,
\label{16}
\end{eqnarray}
\begin{eqnarray}
\left\{\frac{m -\omega \chi}{\sin \theta}+\cal{L}
\right\}F_1+\left\{ \frac{ma-\omega(r^2+ \ell^2
+a^2)}{\Delta^{1/2}}
- \cal{D} \right\}F_2 \nonumber \\
- \mu \left(ir +(\ell+ a \cos \theta) \right)F_4=0,
\label{17}
\end{eqnarray}
where
$$
\mathcal{D}= \frac{i}{2}\left( \frac{r-M}{\Delta^{1/2}} + \Delta^{1/2} \frac{\partial}{\partial
r}\right), \qquad \mathcal{L}=\frac{1}{2} \cot \theta +\frac{\partial}{\partial
\theta}.
$$
Equations (\ref{14}), (\ref{15}), (\ref{16}) and (\ref{17}) imply
the separability ansatz
\begin{eqnarray}
F_1=R_1(r) S_1(\theta),\nonumber\\
F_2=R_2(r) S_2(\theta),\\
F_3=R_2(r) S_1(\theta),\nonumber\\
F_4=R_1(r) S_2(\theta).\nonumber \label{18}
\end{eqnarray}
With the ansatz above, equations take the following forms:
\begin{eqnarray}
\left[\left\{ \frac{ma- \omega (a^2+r^2 + \ell^2)}{\Delta^{1/2}}- \mathcal{D}\right\}R_2(r)-i \mu rR_1(r) \right]
S_1(\theta)\nonumber \\
+\left[\left\{ \frac{m- \omega \chi }{\sin \theta} -
\mathcal{L}\right\}S_2(\theta)+ \mu (\ell +a \cos \theta) S_1
(\theta)  \right]R_1(r)=0, \label{19}
\end{eqnarray}
\begin{eqnarray}
\left[\left\{ \frac{-m + \omega \chi}{\sin \theta}-
\mathcal{L}\right\}S_1(\theta) + \mu(\ell +a \cos
\theta)S_2(\theta) \right]
R_2(r) \nonumber \\
+\left[\left\{ \frac{-ma+ \omega (a^2+r^2 + \ell^2)}{\Delta^{1/2}}
- \mathcal{D}\right\}R_1 (r) -i \mu r R_2(r) \right]S_2(
\theta)=0, \label{20}
\end{eqnarray}
\begin{eqnarray}
\left[\left\{ \frac{ma- \omega (a^2+r^2 + \ell^2)}{\Delta^{1/2}}+
\mathcal{D}\right\}R_1(r)+i \mu rR_2(r) \right]
S_1(\theta)\nonumber \\
+\left[\left\{ \frac{m- \omega \chi }{\sin \theta} -
\mathcal{L}\right\}S_2(\theta)+ \mu (\ell +a \cos \theta) S_1
(\theta)  \right]R_2(r)=0, \label{21}
\end{eqnarray}
\begin{eqnarray}
\left[\left\{ \frac{m-\omega \chi}{\sin \theta}+
\mathcal{L}\right\}S_1(\theta)- \mu(\ell +a \cos
\theta)S_2(\theta) \right]
R_1(r) \nonumber \\
+\left[\left\{ \frac{ma-\omega (a^2+r^2 + \ell^2)}{\Delta^{1/2}} -
\mathcal{D}\right\}R_2 (r) -i \mu r R_1(r) \right]S_2( \theta)=0.
\label{22}
\end{eqnarray}
These equations further imply that
\begin{eqnarray}
\lambda_1 R_1 (r)=\left\{ \frac{ma- \omega (a^2+r^2 + \ell^2)}{\Delta^{1/2}}- \mathcal{D}\right\}R_2(r)- i \mu r R_1(r),\label{23}\\
\lambda_2 R_2 (r)=\left\{ \frac{-ma+ \omega (a^2+r^2 + \ell^2)}{\Delta^{1/2}}- \mathcal{D}\right\}R_1(r)- i \mu r R_2(r),\label{24}\\
\lambda_3 R_2 (r)=\left\{ \frac{ma- \omega (a^2+r^2 + \ell^2)}{\Delta^{1/2}}+ \mathcal{D}\right\}R_1(r)+ i \mu r R_2(r),\label{25}\\
\lambda_4 R_1 (r)=\left\{ \frac{ma- \omega (a^2+r^2 +
\ell^2)}{\Delta^{1/2}}+ \mathcal{D}\right\}R_2(r)- i \mu r
R_1(r)\label{26},
\end{eqnarray}
and
\begin{eqnarray}
\lambda_1 S_1 (\theta)= \left\{ \frac{-m +\omega \chi}{\sin
\theta}+\mathcal{L}\right\} S_2 (\theta)- \mu (\ell+ a \cos
\theta) S_1 (\theta),\label{27}\\
\lambda_2 S_2 (\theta)= \left\{ \frac{m -\omega \chi}{\sin
\theta}+\mathcal{L}\right\} S_1 (\theta)- \mu (\ell+ a \cos
\theta) S_2 (\theta),\label{28}\\
\lambda_3 S_1 (\theta)= \left\{ \frac{-m +\omega \chi}{\sin
\theta}+\mathcal{L}\right\} S_2 (\theta)- \mu (\ell+ a \cos
\theta) S_1 (\theta),\label{29}\\
\lambda_4 S_2 (\theta)= \left\{ \frac{-m +\omega \chi}{\sin
\theta}-\mathcal{L}\right\} S_1 (\theta)+ \mu (\ell+ a \cos
\theta) S_2 (\theta).\label{30}
\end{eqnarray}
For consistency of the equations (\ref{23})-(\ref{30}), we choose
$\lambda_1=\lambda_3=\lambda_4=\lambda$ and $\lambda_2 =-\lambda$.
Then we obtain following independent radial and angular equations:
\begin{eqnarray}
\lambda R_1 (r)=\left\{\frac{ma- \omega (a^2+r^2 + \ell^2)}{\Delta^{1/2}}- \mathcal{D} \right\}R_2 (r) -i \mu r R_1 (r), \label{31}\\
\lambda R_2 (r)=\left\{ \frac{ma- \omega (a^2+r^2 +
\ell^2)}{\Delta^{1/2}}+\mathcal{D}\right\}R_1 (r) +i \mu r R_2
(r),\label{32}
\end{eqnarray}
and
\begin{eqnarray}
\lambda S_2 (\theta)=\left\{ \frac{\omega \chi -m}{\sin \theta} -
\mathcal{L}\right\}S_1 (\theta)+\mu (\ell +a \cos \theta) S_2
(\theta),\label{33}\\
\lambda S_1 (\theta)=\left\{ \frac{\omega \chi -m}{\sin \theta} +
\mathcal{L}\right\}S_2 (\theta)- \mu (\ell +a \cos \theta) S_1
(\theta). \label{34}
\end{eqnarray}

\subsection{Angular equations}

Angular equations (\ref{33}) and (\ref{34}) can be arranged as
\begin{equation}
\frac{dS_1}{d \theta}+ \left\{ (\frac{1}{2}+2 \omega \ell) \cot \theta -a \omega \sin \theta+
\frac{m}{ \sin \theta} \right\}S_1 (\theta)= \left( \mu(  \ell+a \cos \theta)- \lambda \right)S_2 (\theta) ,  \\
\label{35}
\end{equation}
and
\begin{equation}
\frac{dS_2}{d \theta}+ \left\{(\frac{1}{2}-2 \omega \ell) \cot
\theta +a \omega \sin \theta-
\frac{m}{ \sin \theta} \right\}S_2 (\theta)= \left( \mu(  \ell+a \cos \theta)+ \lambda \right)S_1 (\theta).  \\
\label{36}
\end{equation}

At this stage, we affect the transformation
\begin{eqnarray}
S_1= \cos \left(\frac{\theta}{2}\right) T_1 + \sin \left(\frac{\theta}{2}\right) T_2 \nonumber\\
S_2=- \sin \left(\frac{\theta}{2}\right) T_1 + \cos \left(
\frac{\theta}{2}\right) T_2.  \label{37}
\end{eqnarray}
Then taking $x=\cos \theta$ and redefining $T_1=T_+$ and $T_2=T_-$, one can easily see that (\ref{35}) and (\ref{36}) satisfy the
following second order differential equations:
\begin{equation}
(1-x^{2}) \frac{d^{2} T_{\pm}}{d x^{2}} +  M_{\pm} \frac{d
T_{\pm}}{d x} +  N_{\pm} T_{\pm} = 0 \label{38}
\end{equation}
where
\begin{equation}
 M_{\pm} =  \frac{\ell(2 \omega - \mu ) ( 1 - x^{2} ) - 2 a
( \mu - \omega ) x (1- x^{2} )}{\left( \mp \left( \frac{1}{2} +
\lambda \right) - m \right) - \ell ( 2 \omega - \mu ) x + a \omega
+ ( a \mu -  a \omega )x^{2} } - 2 x \label{39}
\end{equation}
and
\begin{eqnarray}
&  &  N_{\pm}= - \frac{\left(m \pm \frac{1}{2} \right)^2 +4
\omega \ell \left(m \pm \frac{1}{2} \right) x +4 \omega^2 \ell^2 }
{ 1 - x^{2} }
\nonumber \\
\nonumber \\
& & + \left( \pm \ell (\mu- 2 \omega )+ (2 \omega^2 - \mu^2)2 \ell
a\right) x + (\pm 2-a \mu -a \omega) (a \mu -a \omega) x^{2}
\nonumber \\
\nonumber \\&  & + (4 \omega^2 -\mu^2) \ell^2 \mp (\mu -\omega)a
-a^2 \omega^2 +2ma \omega +\lambda ( \lambda + 1 ) \label{40} \\
\nonumber \\ & & +  \left( \frac{(\frac{1}{2}\pm m)x \pm 2\omega
\ell \pm \left( \ell(\mu -2\omega )+a(\mu -\omega) x \right)( 1 -
x^{2} )}{ \left(\mp(\frac{1}{2}+\lambda) -m \right)-\ell(2 \omega-
\mu ) x +
a \omega + ( a \mu - a \omega ) x^{2} } \right) \times \nonumber \\
\nonumber \\
& & \left( 2 a ( \mu - \omega) x - (2 \omega - \mu ) \ell \right).
\nonumber
\end{eqnarray}

Now we investigate exact solutions to equations (\ref{38}) for some special
cases:

\noindent {\bf i.}  $ \ell = 0 $, $ a = 0 $ :

In that case, equations (\ref{38}) take the simple form,
\begin{equation}
\frac{d }{ d x} \left( ( 1- x^{2} ) \frac{ d T_{\pm} }{d x}
\right) + \left( \lambda ( \lambda + 1 ) - \frac{ \left( m \pm
\frac{1}{2} \right)^{2} }{ 1 - x^{2} } \right) T_{\pm} = 0 .
\label{41}
\end{equation}
In general for generic values of $\lambda$, the solutions to those equations can be expressed in terms of associated Legendre functions
$P_{\lambda}^{\nu_{\pm}}$ and $Q_{\lambda}^{\nu_{\pm}}$, with $ \nu_{\pm}=m\pm \frac{1}{2}$.
When $\lambda$ and $\nu_\pm$ are integers ($\nu_\pm$ being even), solutions describe associated Legendre polynomials.

\noindent {\bf ii.}  $\ell = 0 $, $ a \neq 0 $ and $\mu= \omega$:

For that special case, equations (\ref{38}) can be simplified to
\begin{equation}
\frac{d }{ d x} \left( ( 1- x^{2} ) \frac{ d T_{\pm} }{d x}
\right) + \left( \widetilde{\lambda} ( \widetilde{\lambda}+ 1 ) -
\frac{ \left( m \pm \frac{1}{2} \right)^{2} }{ 1 - x^{2} } \right)
T_{\pm} = 0 , \label{42}
\end{equation}
where
\begin{equation}
\widetilde{\lambda}(\widetilde{\lambda} +1)=\lambda(\lambda +1)+2m
a \omega -a^2 \omega^2.\label{43}
\end{equation}
Again the solutions are the same as in the case ({\bf i})
except that $\lambda$ is replaced by $\widetilde{\lambda}$.

\noindent{\bf iii.}  $ \ell = 0 $, $ a \neq 0 $, $ \mu \neq \omega$:

In that case, the equations (\ref{38}) can be reduced to
\begin{equation}
(1-x^2)^2 \frac{d^2 T_{\pm} }{ d x^2}+ \left(C x^4 + D_{\pm} x^2 +
E_{\pm} \right) T_{\pm}=0, \label{44}
\end{equation}
where we restrict the eigenvalue
\begin{equation}
\lambda= \pm a \mu \mp m -\frac{1}{2}. \label{45}
\end{equation}
Here
\begin{eqnarray}
C& =&  a^2 \mu^2 - a^2 \omega^2,\nonumber \\
D_{\pm}& = & -a^2 \mu^2 +2 a^2 \omega^2 \mp m \mp a \omega -
\lambda^2
-2a \omega m - \frac{1}{2},\label{46}\\
E_{\pm}& = &- \left(m \pm \frac{1}{2} \right)^2 \pm a \omega
+\lambda^2 +2ma \omega -a^2 \omega^2 -\frac{1}{2}. \nonumber
\end{eqnarray}
Equations of the type (\ref{44}) have exact analytical solutions
for $C=0$, $D_{\pm} =0$ or $C=0$, $D_{\pm} \neq 0$ \cite{polyanin}.
However in our case since $\mu \neq \omega$, $C\neq0$. In fact,
when $\lambda$ is further restricted to be $\lambda=\pm \mu a$ (which corresponds to $m=\mp \frac{1}{2}$),
equations (\ref{44}) have simple analytical  solutions. When the mass of the particle is greater than the frequency of the
spinor wave ($\mu > \omega$), the solutions describe periodic waves. However if the mass of the particle is smaller than the frequency of the
spinor wave ($\mu < \omega$), the solutions are exponential.

\noindent {\bf iv.}  $\ell \neq 0 $, $ a = 0 $ and $\mu = 2 \omega $:

In that case, equations (\ref{38}) can be arranged as
\begin{eqnarray}
(x^{2}-1)^2 \frac{d^{2} T_{\pm}}{d x^{2}}+2x(x^2-1) \frac{d
T_{\pm}}{d x}+ \left(C_{\pm}+D_{\pm} x - \lambda(\lambda+1)x^2
\right)T_{\pm}=0,\label{47}
\end{eqnarray}
where
\[
C_{\pm}=\lambda(\lambda +1)-\left( m \pm \frac{1}{2}\right)^2 -4
\omega^2 \ell^2, \qquad D_{\pm}=-4 \omega \ell \left( m \pm
\frac{1}{2} \right).
\]
Under the transformation
\begin{equation}\xi=\frac{1}{2}(1-x),
\qquad Y_{\pm}=(x+1)^{-p_{\pm}} (1-x)^{-q_{\pm}} T_{\pm} \label{48}
\end{equation}
the equation satisfied by $Y_{\pm}$ takes the form
\begin{equation}
\xi(\xi-1) \frac{d^2 Y_{\pm}}{d \xi^2}+[\xi (\alpha_{\pm} +\beta_{\pm}
+1)-\gamma_{\pm}] \frac{d Y_{\pm}}{d \xi}+\alpha_{\pm}\, \beta_{\pm}\, Y_{\pm} =0 \label{49}
\end{equation}
whose solution is given by hypergeometric function
\begin{equation}
Y_{\pm}(\xi)=F(\alpha_{\pm}, \beta_{\pm},\gamma_{\pm}; \xi).
 \label{50}
\end{equation}
Hence the solutions are
\begin{equation}
T_{\pm}(x)=(1+x)^{p_{\pm}}(1-x)^{q_{\pm}}F \left(\alpha_{\pm} ,\beta_{\pm}, \gamma_{\pm};\frac{1}{2} (1-x) \right).\label{51}
\end{equation}
Here
\begin{equation}
q_{\pm}^2=\frac{1}{4} \left( m \pm \frac{1}{2}+2 \omega \ell \right)^2,\label{52}
\end{equation}
\begin{equation}
p_{\pm}^2=\frac{1}{4} \left( m \pm \frac{1}{2}-2 \omega \ell \right)^2.\label{53}
\end{equation}
$\alpha_{\pm}, \beta_{\pm}$ and $\gamma_{\pm}$ can be obtained from
\begin{eqnarray}
\alpha_{\pm} +\beta_{\pm} +1&=& 2(p_{\pm}+q_{\pm}+1), \nonumber\\
\alpha_{\pm} \beta_{\pm} &=& (p_{\pm} +q_{\pm})^2 -  p_{\pm}- q_{\pm}+ 2(p_{\pm} +q_{\pm})
-\lambda(\lambda+1),\label{54}\\
\gamma_{\pm}&=& 2q_{\pm} +1. \nonumber
\end{eqnarray}

\noindent {\bf v.} $\ell \neq 0 $, $ a = 0 $ and $\mu \neq2 \omega$:

In that case, equations (\ref{38}) reduce to
\begin{eqnarray}
& & (1-x^2) \frac{d^2 T_{\pm} }{ d x^2}+ \left(-2x + \frac{\ell (2
\omega - \mu ) (1-x^{2})}{\left(\mp(\frac{1}{2} + \lambda )-m
\right)- \ell( 2 \omega - \mu ) x } \right) \frac{d T_{\pm} }{ d
x} \nonumber\\
& & + \left( - \frac{(m\pm \frac{1}{2})^2 + 4 \omega \ell \left( m
\pm \frac{1}{2} \right) x + 4 \omega^{2} \ell^{2}}{1-x^2 } \pm
\ell ( \mu - 2 \omega ) x  + \lambda ( \lambda + 1 )
\right.\label{55} \\ & & \left. + ( 4 \omega^{2} - \mu^{2} )
\ell^{2}  + \frac{( \frac{1}{2} \pm m)x \pm 2 \omega \ell \pm \ell
( \mu - 2 \omega ) (1-x^2)}{ \left( \mp( \frac{1}{2} + \lambda )-m
\right) - \ell ( 2 \omega - \mu ) x } ( \mu - 2 \omega ) \ell
\right)T_{\pm}=0. \nonumber
\end{eqnarray}
As we have done in the case {\bf iii.}, we simplify (\ref{55}) by
taking the constraints
\begin{equation}
\mu \ell -2 \omega \ell= \left(\frac{1}{2} +\lambda \right) \pm m \label{56}
\end{equation}
in the equations satisfied by $T_+$ and $T_-$ respectively. With
these restrictions, equations (\ref{55}) satisfied for $T_+$ and $T_-$ reduce to
\begin{equation}
(1-x^2)^2 \frac{d^2 T_{+} }{ d x^2}+(1-x)(1-x^2)\frac{d T_{+}
}{ d x} + \left(\bar{C}_{+} x^2 + \bar{D}_{+} x + \bar{E}_{+}
\right) T_{+}=0, \label{57}
\end{equation}
and
\begin{equation}
(1-x^2)^2 \frac{d^2 T_{-} }{ d x^2}-(1+x)(1-x^2)\frac{d T_{-}
}{ d x} + \left(\bar{C}_{-} x^2 + \bar{D}_{-} x + \bar{E}_{-}
\right) T_{-}=0, \label{58}
\end{equation}
respectively, where
\begin{eqnarray}
\bar{C}_{\pm}& =& \ell (\mu -2 \omega)(4 \omega \ell +2 m \pm 1)-(m \pm \frac{1}{2})^2,\nonumber\\
\bar{D}_{\pm}& = & -4 \omega \ell (m \pm 1)-\left(m \pm \frac{1}{2} \right),\label{59} \\
\bar{E}_{\pm}& = & 4 \omega \ell ( \omega \ell -\mu \ell \pm m)-\left(\frac{1}{2}\pm m \right)(2 \mu \ell +1).\nonumber
\end{eqnarray}
Equations of the type (\ref{57}) and (\ref{58}) seem harder to obtain for exact analytical solutions. However, with
$\lambda= \mu \ell$ both equations take the following simple forms:
\begin{equation}
(1+x)^2 \frac{d^2 T_{+} }{ d x^2}+(1+x)\frac{d T_{+}
}{ d x} -\left(m+\frac{1}{2} \right)^2T_{+}=0\label{60}
\end{equation}
and
\begin{equation}
(x-1)^2 \frac{d^2 T_{-} }{ d x^2}+(x-1)\frac{d T_{-}
}{ d x} -\left(m-\frac{1}{2} \right)^2T_{-}=0,\label{61}
\end{equation}
whose solutions are given by
\begin{equation}
T_{+}(x)=c_1 (1+x)^{(m+ \frac{1}{2})}+c_2 (1+x)^{-(m+\frac{1}{2})},\label{62}
\end{equation}
\begin{equation}
T_{-}(x)=d_1(1-x)^{(m- \frac{1}{2})}+d_2 (1-x)^{-(m- \frac{1}{2})},\label{63}
\end{equation}
where $c_1$, $c_2$, $d_1$ and  $d_2$ are real  constants.

\vspace{0.8cm}
\noindent {\bf vi.} $\ell \neq 0 $, $ a \neq 0 $ and $\mu = 2 \omega$:

In that case, equations (\ref{38}) take the following form:
\begin{eqnarray}
& & (1-x^2) \frac{d^2 T_{\pm} }{ d x^2}- \left( 2x + \frac{2a \omega x(1-x^2)}{\left(\mp(\frac{1}{2} + \lambda )-m
\right)+a \omega+a \omega  x^2 } \right) \frac{d T_{\pm} }{ d
x} \nonumber\\
& & + \left( - \frac{(m\pm \frac{1}{2})^2 + 4 \omega \ell \left( m
\pm \frac{1}{2} \right) x + 4 \omega^{2} \ell^{2}}{1-x^2 }
 + \lambda ( \lambda + 1 )   \right. \label{64}\\
& &  \left. - 4 \ell a \omega^2 x +(\pm2 -3 a\omega)a \omega x^2
  \mp a \omega -a^2 \omega^2 +2ma \omega \right.\nonumber\\
 & &  \left. + 2a \omega x \frac{(\frac{1}{2}\pm m)x \pm 2 \omega \ell \pm a \omega x (1-x^2)}{\left(\mp(\frac{1}{2} + \lambda )-m
\right)+a \omega+a \omega  x^2}\right)T_{\pm}=0.\nonumber
\end{eqnarray}
Under the constraints
\begin{equation}
2a \omega=m \pm \left(\frac{1}{2}+ \lambda \right), \label{65}
\end{equation}
equations (\ref{64}) can be simplified as
\begin{equation}
(1-x^2)^2 \frac{d^2 T_{+} }{ d x^2}+\left( a_0 + a_1 x +a_2 x^2 +a_3 x^3 +a_4 x^4 \right)T_{+}=0,\label{66}
\end{equation}
and
\begin{equation}
(1-x^2)^2 \frac{d^2 T_{-} }{ d x^2}+\left(\bar{a}_0 + \bar{a}_1 x +\bar{a}_2 x^2 +\bar{a}_3 x^3 +\bar{a}_4 x^4 \right)T_{-}=0,\label{67}
\end{equation}
respectively, where
\begin{eqnarray}
a_0&=&3 a^2 \omega^2 -4 \omega^2 \ell^2 - \left( m+ \frac{1}{2}\right)\left( 1+ 2a \omega\right), \nonumber \\
a_1&=& -4 \ell \omega \left(m+\frac{3}{2}+ a \omega  \right),\nonumber\\
a_2&=& -6 a^2 \omega^2 + \left(m+\frac{1}{2} \right)\left(2a \omega -m - \frac{3}{2} \right),\label{68}\\
a_3&=&4 a \ell \omega^2, \nonumber\\
a_4&=&3 a^2 \omega^2, \nonumber
\end{eqnarray}
and
\begin{eqnarray}
\bar{a}_0&=&3 a^2 \omega^2 -4 \omega^2 \ell^2 +\left( m- \frac{1}{2}\right) \left(1-2a \omega \right), \nonumber \\
\bar{a}_1&=& -4 \ell \omega \left(m+ a \omega -\frac{3}{2} \right),\nonumber\\
\bar{a}_2&=& -6 a^2 \omega^2 + \left(m - \frac{1}{2} \right) \left( 2 a \omega -m+ \frac{3}{2} \right),\label{69}\\
\bar{a}_3&=&4 a \ell \omega^2,\nonumber \\
\bar{a}_4&=&3 a^2 \omega^2. \nonumber
\end{eqnarray}
In order to obtain an exact analytical solution, equation (\ref{66}) can further be simplified when $\lambda=2 a \omega +1$ (which corresponds to $m=-\frac{3}{2}$)  and $ \ell = \frac{1}{2 \omega}$ as
\begin{equation}
(1+x) \frac{d^2 T_{+} }{ d x^2}+\left( 3 a^2 \omega^2 +2 a \omega + 3a^2 \omega^2 x \right)T_{+}=0.\label{70}
\end{equation}
Then the solution for $T_+$ can be written as
\begin{equation}
T_+(x)= e^{i \sqrt{3}\omega a x}z \Phi(\bar{a}+1, 2, z), \label{71}
\end{equation}
with $z=-2i \sqrt{3}\omega a (x+1)$ and $\bar{a}=-\frac{i}{\sqrt{3}}$.
Similarly (\ref{67}) take a simpler form when $\lambda=2 a \omega - 1$ (which corresponds to $m=\frac{3}{2}$)  and $ \ell = \frac{1}{2 \omega}$ as
\begin{equation}
(1-x) \frac{d^2 T_{-} }{ d x^2}+\left( 3 a^2 \omega^2 -2 a \omega - 3a^2 \omega^2 x \right)T_{-}=0,\label{72}
\end{equation}
whose solution can be given as
\begin{equation}
T_-(x)= e^{-i \sqrt{3}\omega a x}\zeta \Phi(\bar{a}+1, 2, \zeta) \label{73}
\end{equation}
with $\zeta=-2i \sqrt{3}\omega a (x-1)$. For both $T_+ $ and $T_- $, the solutions describe oscillating wave solutions with non-uniform amplitudes.

\subsection{Radial equations}
Radial equations (\ref{36}), (\ref{37}) can be rearranged as
\begin{equation}
\frac{dR_1}{dr} + \left(\frac{r-M}{\Delta} -i \frac{\left ( ma - \omega(a^2 +r^2 + \ell^2)\right)2}{\Delta} \right) R_1= - \frac{2}{\sqrt{\Delta}} \left( \mu r +i \lambda \right) R_2,\label{97}
\end{equation}
\begin{equation}
\frac{dR_2}{dr} + \left(\frac{r-M}{\Delta} +i \frac{\left ( ma - \omega(a^2 +r^2 + \ell^2)\right)2}{\Delta} \right) R_2= - \frac{2}{\sqrt{\Delta}} \left( \mu r -i \lambda \right) R_1 .\label{98}
\end{equation}
To get the radial equations in the form of a wave equation, we follow the method applied in Chandrasekhar's book \cite{chandrasekhar2} but with $\ell \neq 0$. Hence we consider the transformations
\begin{equation}
P_1=i \Delta^{1/2} R_1, \qquad
P_2= \Delta^{1/2} R_2. \label{104}
\end{equation}
Let $\Omega=r^2+a^2+ \ell^2-\frac{ma}{\omega}$.
With these transformations, equations (\ref{97}) and (\ref{98}) take the form
\begin{eqnarray}
\frac{dP_1}{dr}+ 2i \frac{\omega \Omega}{\Delta}P_1= \frac{-2i (\mu r +i \lambda)}{\Delta^{1/2}} P_2, \label{105}\\
\frac{dP_2}{dr}- 2i \frac{\omega \Omega}{\Delta}P_2= \frac{2i (\mu r -i \lambda)}{\Delta^{1/2}} P_1. \label{106}
\end{eqnarray}
Let
\begin{equation}
\frac{du}{dr}=\frac{\Omega}{\Delta}, \qquad \beta^2=M^2 -a^2 + \ell^2. \label{107}
\end{equation}
Then in terms of the new independent variable $u $, we obtain
\begin{eqnarray}
\frac{dP_1}{du}+2i \omega P_1 = \frac{-2i (\mu r +i \lambda)}{\Omega} \Delta^{1/2} P_2, \label{108}\\
\frac{dP_2}{du}-2i \omega P_2 = \frac{2i (\mu r -i \lambda)}{\Omega} \Delta^{1/2} P_1, \label{109}
\end{eqnarray}
where
\begin{eqnarray}
u&=&r+\frac{2 M r_+ +2 \ell^2 -\frac{ma}{\omega}}{2 \beta} \ln \left(\frac{r-r_+}{r_+} \right) \nonumber\\
& & - \frac{2 M r_- +2 \ell^2 -\frac{ma}{\omega}}{2 \beta} \ln \left(\frac{r-r_-}{r_-} \right),\qquad (r>r_+) \label{110}
\end{eqnarray}
for $M^2 > a^2 - \ell^2$. Here $ r_+=M+ \beta$ and $ r_-=M-\beta$.
The relation (\ref{110}) is single-valued for $r>r_+$  if the inequality
\begin{equation}
r_+ ^2 +a^2 +\ell^2 -\frac{ma}{\omega}>0 \label{110-a}
\end{equation}
is satisfied. This requires that in the frequency range
\begin{equation}
\omega_2 < \omega < \omega_1,
\label{110-b}
\end{equation}
where
\begin{equation}
\omega_2=\frac{ma}{2Mr_+ +2 \ell^2}, \qquad  \omega_1 =\frac{ma}{\ell^2 +a^2}, \qquad m>0, \label{110-c}
\end{equation}
$u$ becomes single-valued  in $r$ in the region where $r> r_+$. It is seen that as  $r \rightarrow \infty$, $u \rightarrow \infty$ and as $r \rightarrow (r_+)^+$, $u \rightarrow -\infty$.
For completeness, we also note that for the critical case when $M^2 = a^2 - \ell^2$, $u$ becomes
\begin{eqnarray}
u=r- \frac{\left(2 \ell^2 +2 M^2 -\frac{ma}{\omega} \right)}{r-M}+ 2M \ln(r-M), \qquad (r>M).\label{111}
\end{eqnarray}
Concentrating on the case $M^2 > a^2 - \ell^2$, let us make another transformation
\begin{equation}
\vartheta= \arctan\left( {\frac{\mu r}{\lambda}}\right). \label{112}
\end{equation}
With the new definitions
\begin{equation}
P_1=\phi_1 e^{-\frac{1}{2}i\vartheta }, \qquad P_2=\phi_2 e^{\frac{1}{2}i\vartheta },  \label{113}
\end{equation}
equations (\ref{108}), (\ref{109}) take the forms
\begin{eqnarray}
\frac{d\phi_1}{du}+ 2i\omega \left(1- \frac{\lambda \mu \Delta}{4 \omega (\lambda^2+ \mu^2 r^2)\Omega} \right)\phi_1=2 \frac{\sqrt{\lambda^2+ \mu^2 r^2}}{\Omega}\Delta^{1/2}\phi_2,  \label{114}\\
\frac{d\phi_2}{du}- 2i\omega \left(1- \frac{\lambda \mu \Delta}{4 \omega (\lambda^2+ \mu^2 r^2)\Omega} \right)\phi_2=2 \frac{\sqrt{\lambda^2+ \mu^2 r^2}}{\Omega}\Delta^{1/2}\phi_1.  \label{115}
\end{eqnarray}
Redefining the independent variable as
\begin{equation}
\hat{u}=u-\frac{1}{4\omega}\arctan \left({\frac{\mu r}{\lambda}} \right),\label{116}
\end{equation}
equations (\ref{114}) and (\ref{115}) can be simplified as
\begin{eqnarray}
\frac{d\phi_1}{d\hat{u}}+2i\omega \phi_1 =W\phi_2, \label{117}\\
\frac{d\phi_2}{d\hat{u}}-2i\omega \phi_2 =W\phi_1, \label{118}
\end{eqnarray}
where
\begin{equation}
W=\frac{2\left(\lambda^2+ \mu^2 r^2 \right)^{3/2} \Delta^{1/2}}{\left(\lambda^2+ \mu^2 r^2 \right)\Omega -\frac{\lambda \mu \Delta}{4 \omega}}.\label{119}
\end{equation}
By further defining $Z_1=\phi_1 +\phi_2 $ and $Z_2=\phi_1 -\phi_2 $, equations (\ref{117}) and (\ref{118}) can be rewritten
\begin{eqnarray}
\frac{d Z_1}{d\hat{u}}-W Z_1 = -2i \omega Z_2, \label{120}\\
\frac{d Z_2}{d\hat{u}}+W Z_2 = -2i \omega Z_1. \label{121}
\end{eqnarray}
From equations (\ref{120}) and (\ref{121}) we obtain one-dimensional wave equations
\begin{eqnarray}
\frac{d^2 Z_1}{d\hat{u}^2}+4\omega^2 Z_1 =V_+ Z_1, \label{122}\\
\frac{d^2 Z_2}{d\hat{u}^2}+4\omega^2 Z_2 =V_- Z_2, \label{123}
\end{eqnarray}
where the effective potentials
\begin{equation}
V_{\pm}=W^2 \pm \frac{dW}{d\hat{u}}.\label{124}
\end{equation}
We calculate the potentials as
\begin{eqnarray}
V_{\pm} (r)=\frac{2\left(\lambda^2 +\mu^2 r^2 \right)^{3/2} \Delta^{1/2}}{I^2}\left[ 2 \left(\lambda^2 +\mu^2 r^2 \right)^{3/2} \Delta^{1/2}
 \right. \nonumber\\
\pm \left.  3 \mu^2 r \Delta   \pm \left(\lambda^2 +\mu^2 r^2  \right)(r-M) \right.\label{125}\\
\mp\left. \frac{\Delta \left(\lambda^2 +\mu^2 r^2 \right)}{I}\left( 2 \mu^2 \Omega r+ 2\left(\lambda^2 +\mu^2 r^2 \right)r-\frac{\lambda \mu(r-M)}{2 \omega} \right)   \right],\nonumber
\end{eqnarray}
where
\begin{equation}
I=\left(\lambda^2 +\mu^2 r^2 \right) \Omega - \frac{\lambda \mu \Delta}{4 \omega}. \label{126}
\end{equation}
We see that, the effective potentials depend on gravitomagnetic monopole moment $\ell$ via the functions $\Delta$ and $\Omega$, where $\ell=0$ case is discussed in \cite{chandrasekhar2}. We also report that, for $\mu=0$, the potentials take the simple form
\begin{eqnarray}
V_{\pm} (r)=\frac{2  \Delta^{1/2} \lambda }{\Omega^2} \left\{ 2 \lambda \Delta^{1/2} \pm (r-M) \mp \frac{ 2  \Delta r }{\Omega} \right\}. \label{127}
\end{eqnarray}
\section{Discussion}
To see the asymptotic behaviour of the potentials and the radial solutions and to expose the effect of gravitomagnetic monopole moment, we can expand the potentials up to order $ \mathcal{O}( \frac{1}{r^{3}})$.  At this order, for massive case (i.e for $\mu \neq 0 $), the potentials behave as
\begin{equation}
V_{\pm}  \simeq 4 \mu^{2} - 8 \mu^{2} M \frac{1}{r}
 + \eta_{\pm} \frac{1}{r^{2}}+ \mathcal{O}(\frac{1}{r^3}), \label{e1}
\end{equation}
where
\begin{equation}
\eta_{\pm}= \frac{2}{\omega} ( \lambda \mu + 4 \mu^{2} m a + 2 \lambda^{2} \omega - 2 \mu^{2} a^{2} \omega - 6 \mu^{2} \ell^{2} \omega \pm M \mu \omega ).\label{e2}
\end{equation}
Here the first term corresponds to constant value of the potential at the asymptotic infinity. The second term represents the monopole-type (or Coulomb type) potential while the third
term exhibits a dipole-type potential. As can be seen from the asymptotic expansion of the potentials, the effect of NUT parameter is observed at $\frac{1}{r^2}$ order. It means that the effect of NUT charge in the massive case appears in a dipole-type potential at the leading order. In the massless case ($\mu = 0$), the potentials simply take the form
\begin{equation}
V_{\pm} \simeq 4 \lambda^{2} \frac{1}{r^{2}}
\label{e3}
\end{equation}
up to order $ \mathcal{O}( \frac{1}{r^{3}})$.  With the asymptotic form of the potentials given above, radial equations (\ref{122}) and (\ref{123}) take the following forms:
\begin{equation}
\frac{d^2 Z_1}{dr^2}+4\omega^2 Z_1 =\left(4 \mu^{2} - 8 \mu^{2} M \frac{1}{r}
 + \eta_{+} \frac{1}{r^{2}}\right) Z_1, \label{122-1}
\end{equation}
\begin{equation}
\frac{d^2 Z_2}{dr^2}+4\omega^2 Z_2 =\left(4 \mu^{2} - 8 \mu^{2} M \frac{1}{r}
 + \eta_{-} \frac{1}{r^{2}} \right) Z_2, \label{123-1}
\end{equation}
whose solutions can be given by
\begin{equation}
Z_{1,2}=r^{s_{\pm}} e^{2i\sqrt{\omega^2 -\mu^2}r}\Phi (\bar{c}_{\pm}, \bar{d}_{\pm};\xi),\label{e4}
\end{equation}
in terms of confluent hypergeometric functions $ \Phi (\bar{c}_{\pm}, \bar{d}_{\pm};\xi)$, where
$+$ corresponds to solution for $Z_1$, while $-$ corresponds to solution for $Z_2$. We also consider that $\omega > \mu$. Here
\begin{equation}
s_{\pm}=\frac{1+\sqrt{1+4 \eta_{\pm}}}{2} \label{e5}
\end{equation}
and
\begin{equation}
\xi=-4i \sqrt{\omega^2-\mu^2} r, \qquad \bar{c}_{\pm}=s_{\pm}-\frac{2i \mu^2 M}{\sqrt{\omega^2 -\mu^2}}, \qquad \bar{d}_{\pm}=2s_{\pm}. \label{e6}
\end{equation}
For the physically acceptable solutions, the inequality
\begin{equation}
1+4 \eta_{\pm} \geq 0 \label{e7}
\end{equation}
should also be imposed.
At the asymptotic infinity ($r \rightarrow \infty$), the behavior of the solutions (\ref{e4}) can be represented as
\begin{eqnarray}
Z_{1,2} &\sim&
a_1 e^{ip(r)}+a_2 e^{-ip(r)}+(b_1 e^{ip(r)}+b_2 e^{-ip(r)})\frac{1}{r} \label{e8}\\
& & +\left(c_1 e^{i(p(r)+\frac{\pi}{2})}+c_2 e^{i(\frac{\pi}{2}-p(r))}\right)\frac{1}{r^2}+\mathcal{O}(\frac{1}{r^3}),\nonumber
\end{eqnarray}
where
\begin{equation}
p(r) =\frac{\pi}{2} s_{\pm}- 2 \left( \sqrt{\omega^2 - \mu^2} r + \frac{\mu^2 M}{\sqrt{\omega^2 - \mu^2}} \ln(4 \sqrt{\omega^2 - \mu^2} r ) \right),\label{e9}
\end{equation}
$a_1,a_2,b_1,b_2, c_1 $ and $c_2$ are constants coming from the asymptotic expansion of $\Phi (\bar{c}_{\pm}, \bar{d}_{\pm};\xi)$. The asymptotic behaviour (\ref{e8}) represents
incident and reflected planar-type waves plus incident and reflected spherical-type waves at infinity.

Interestingly, in the critical case when $\omega=\mu$, radial wave equations accept the solution
\begin{equation}
Z_{1,2}=r^{1/2} J_{\nu_{\pm}}(\beta \sqrt{r}), \label{e10}
\end{equation}
where $\nu_{\pm}=1-2s_{\pm}$ and $\beta=4\mu \sqrt{2M}$ and $J_{\nu} (\beta \sqrt{r})$ represents Bessel functions of order $\nu$.
\begin{figure}
\includegraphics{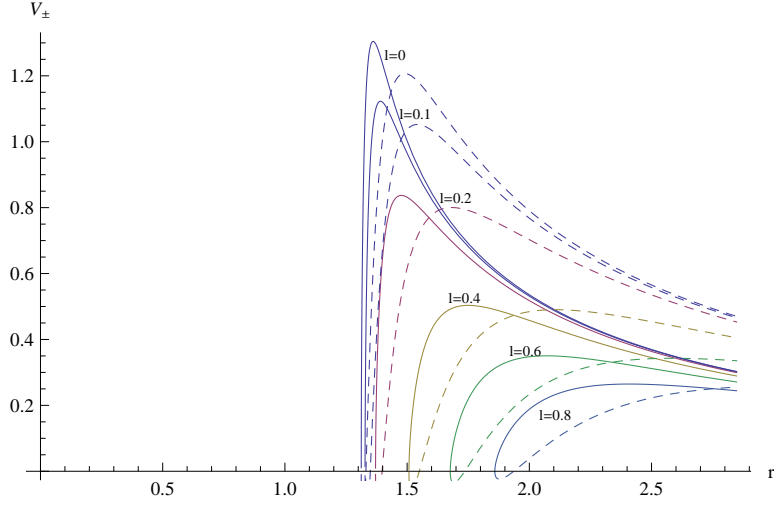}
\caption{ Graphs of $V_+$ (solid curves) and $V_-$ (dashed curves) with $M=1$, $\lambda=1$, $\mu=0.12$,
$\omega=0.2$, $m=0.5$, $a=0.95$} \label{figure1}
\end{figure}

\begin{figure}
\includegraphics{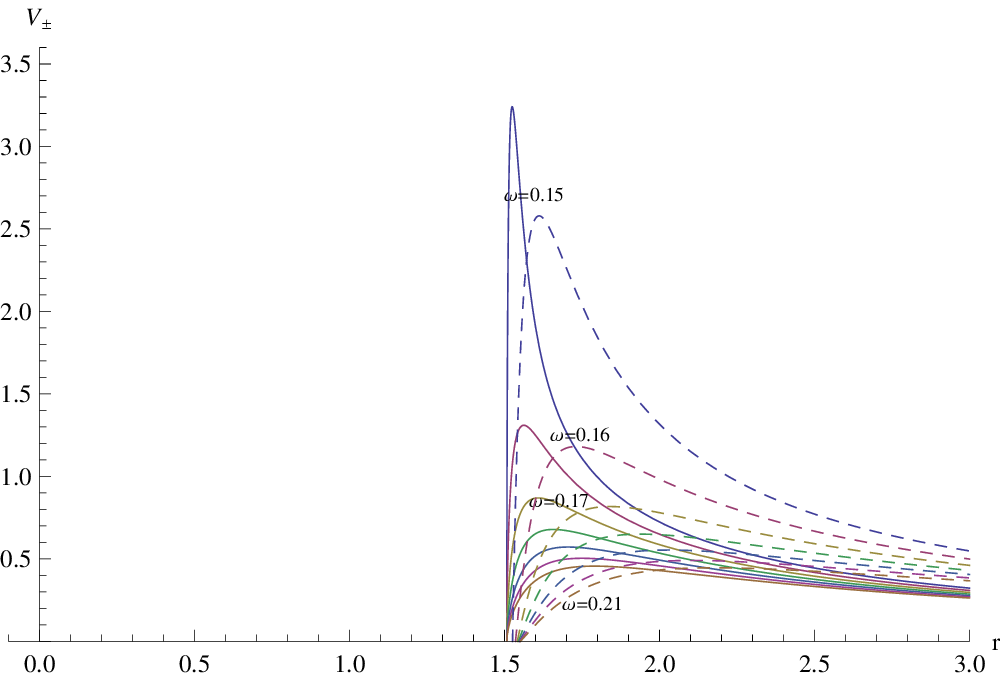}
\caption{ Graphs of $V_+$ (solid curves) and $V_-$ (dashed curves) with $M=1$, $\lambda=1$, $\mu=0.12$,
$\ell=0.4$, $m=0.5$, $a=0.95$}\label{figure2}
\end{figure}



\begin{figure}
\includegraphics{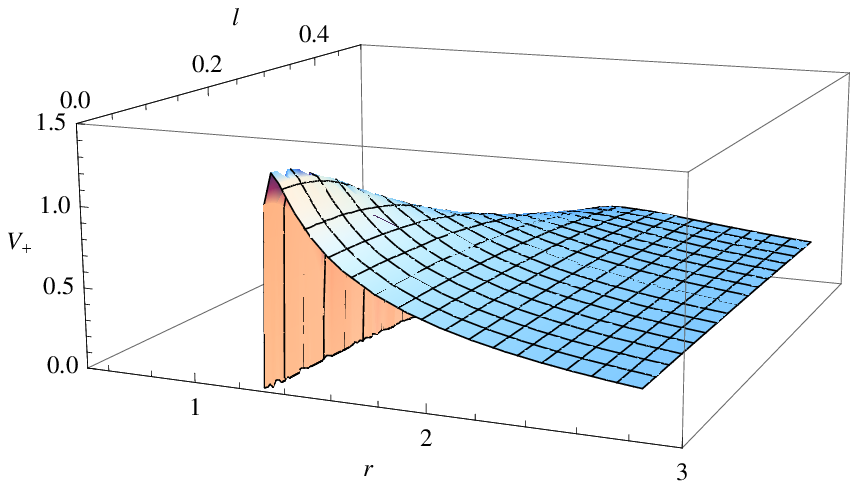}
\caption{3-dimensional plot of $V_+$  with $M=1$, $\lambda=1$, $\mu=0.12$,
$\omega=0.2$, $m=0.5$, $a=0.95$} \label{figure3}
\end{figure}

\begin{figure}
\includegraphics{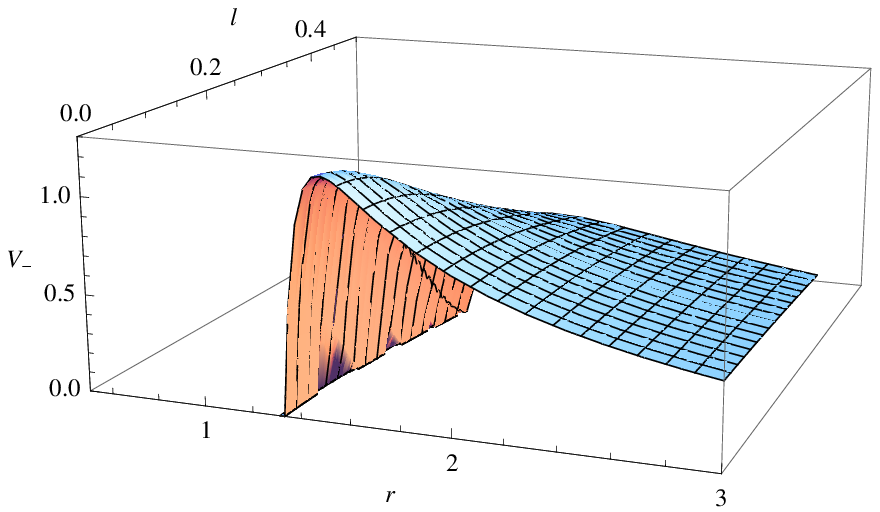}
\caption{3-dimensional plot of $V_-$   with $M=1$, $\lambda=1$, $\mu=0.12$,
$\omega=0.2$, $m=0.5$, $a=0.95$} \label{figure4}
\end{figure}



It would also be interesting to see the behaviour of the potentials graphically. From the expressions (\ref{125}) and (\ref{127}), it is obvious that, the potentials become singular in the  massive case ($\mu \neq 0$) when $I=0$ and in the massless case ($\mu= 0$) when $\Omega=0$. Moreover they possess local extremums when $\frac{dV_{\pm}}{dr}=0$ that leads to a very complicated algebraic equation to solve for the extremum distance of $r$. However, in order to understand the physical behaviour of the potentials $V_{\pm}$ in the physical region $ r > r_{+}$, we make 2-dimensional and 3-dimensional plots of the potentials for massive particles. It is also clear that the potentials depend on the physical parameters $a$, $M$, $m$ and $\ell$ implicitly in the metric functions.
In all plots, we take the physical  parameters $M=1$, $\lambda=1$, $m=0.5$ and $a=0.95$.

As can be seen in 2-dimensional graphs, the potentials are plotted as a function of radial distance $r$.
Figure \ref{figure1} and \ref{figure2} describe the effective potentials $ V_{\pm}$ for massive particle with rest mass $\mu=0.12$ such that $\mu < \omega $. In the first graph, also taking the frequency constant ($\omega=0.2$), we examine the effect of the NUT parameter $ \ell $ by obtaining potential curves for some specific values of the gravitomagnetic monopole moment $\ell$. We see that, for sufficiently small values of $\ell$ including $\ell=0 $, potentials have sharp peaks in the physical region  $r>r_+$. When the NUT parameter $ \ell=0 $, the peak is seen to be maximum. It is also observed that while $\ell$ increases, the sharpness of the peaks decreases. While the peaks get smaller, potentials still have some maxima. The peaks tend to disappear after a specific value of the NUT parameter. This means that, for small values of the NUT parameter, a massive Dirac particle moving in the region $ r_{+} < r < \infty $ may encounter sharp potential barriers resulting in decrease of its kinetic energy, but for sufficiently large values of the gravitomagnetic monopole moment, it may advance in the same region even without encountering any peaks. Potentials become bounded regardless of
the value of $\ell$ and approach a constant value in the sufficiently large values of $r$ (or $r \rightarrow \infty$). In the second graph, we keep gravitomagnetic
monopole moment $\ell$ fixed ($\ell=0.4$). In that case, we investigate the behaviour of the potentials by obtaining potential curves for some specific values of the frequency $\omega$ that can take values in the range
(\ref{110-b}) for $\hat{u}$ or $u$ to be single-valued. Again, one can clearly see that potentials have some local maxima in the low frequencies and the peaks are observed. While the frequency increases, the peaks again disappear as in Figure \ref{figure1} and potentials behave similarly in the sufficiently large distances.
We also remark that, in the massless case ($\mu=0$), 2-dimensional plots of the behaviour of the potentials are similar to figures \ref{figure1} and \ref{figure2}.

To observe the effect of the NUT parameter $\ell$ explicitly, we also realize 3-dimensional plots of the potentials with respect to NUT parameter $\ell$ and radial distance $r$. As can be seen from the 3-dimensional graphs \ref{figure3} and \ref{figure4}, we observe 3-dimensional peak for small values of gravitomagnetic moment. As the value of NUT parameter and radial distances increases, potentials level off. Again in the massless case, 3-dimensional plots of the behaviour of the potentials are similar to figures \ref{figure3} and \ref{figure4}.
\section{Conclusion}
In this work, we examine the Dirac equation in 4-dimensional Kerr-Taub-NUT spacetime described by the physical parameters, the mass $M$, angular speed $a$ per unit mass and
gravitomagnetic monopole moment $\ell$. By taking an axially symmetric ansatz for the spinor field, we obtain massive Dirac equations. By using Boyer-Lindquist coordinates, we explicitly work out the separability of the equations into radial and angular parts. We get angular and radial equations for arbitrary $\ell$. We find some exact solutions to the angular equations with and without gravitomagnetic monopole moment $\ell$ and rotation parameter $a$. We see that, for massive Dirac equation, when the mass of the particle is equal to or twice the frequency of the spinor wave function, some angular solutions can be represented in terms of hypergeometric functions.

We also discuss the radial equations and get a wave equation with an effective potential. We obtain asymptotic expansion of the potentials to observe the effect of the NUT parameter. We have seen that the effect of NUT charge manifests itself  in a dipole-type potential at the leading order. With the asymptotic form of the potentials, we get the solutions of the radial wave equations. We have seen that, the radial wave functions physically represent plane wave-type and spherical wave-type solutions. Moreover, to realize the physical interpretations of the potentials, we make 2 and 3 dimensional plots of them. From the plots, it can be seen that the peak values of the potential barriers decrease and the potential curves level off while NUT charge $\ell$ increases.

We believe that in order to better understand the physical significance of the NUT charge, Dirac Hamiltonian should be constructed for the Dirac equation \cite{dereli1} and e.g. the effect of NUT parameter on the neutrino oscillations can be examined. As a future work, it can be further suggested that, by using a similar spectral method presented in \cite{dolan}, angular solutions can be represented in terms of spheroidal harmonics and eigenvalues can be solved numerically. Finally, we remark that one can study the massive Dirac equation for a charged Dirac particle in the background of Kerr-Newman-Taub NUT spacetime as well. These are devoted to future research.

\section*{\label{ackno} Acknowledgments}
We would like to thank the anonymous referee whose useful suggestions and comments led us to improve our manuscript.

\section*{Appendix A: Derivation of the solutions of (\ref{70}) and (\ref{72})}

Equations (\ref{70}) and (\ref{72}) are of the type
\begin{equation}
(c_2 x+b_2)\frac{d^2 y }{ d x^2}+(c_1 x+b_1)\frac{d y }{ d x}+(c_0 x+b_0)y =0, \label{A1}
\end{equation}
with $c_1=0$, $b_1=0$. Under the transformation \cite{polyanin}
\begin{equation}
y=e^{kx} \Omega(z), \qquad z=\frac{1}{\Lambda}(x- \bar{\mu})\label{A2}
\end{equation}
equation satisfied by $ \Omega(z)$ takes the form
\begin{equation}
z \frac{d^2 \Omega }{ d z^2}+ (\bar{b} -z)\frac{d \Omega }{ d z}-\bar{a} \Omega=0, \label{A3}
\end{equation}
where
\begin{equation}
\bar{a}=\frac{b_2 k^2 +b_0}{2 c_2 k}, \qquad \bar{b}=0. \label{A4}
\end{equation}
Here $\bar \mu=- \frac{b_2}{c_2}$ and  $\Lambda=- \frac{1}{2k}$.
$k$ can be calculated from $c_2 k^2 +c_0 =0$. A particular solution of \ref{A3} (with $\bar{b}=0$) can be written in terms of
confluent hypergeometric function as
\begin{equation}
\Omega(z)=z \Phi(\bar{a}+1, 2, z). \label{A5}
\end{equation}
In our case, for $T_+$
\begin{equation}
c_2=1, \qquad b_2=1, \qquad c_0= 3a^2 \omega^2, \qquad b_0=3a^2 \omega^2+ 2a \omega. \nonumber
\end{equation}
So the solution for $T_+$ can be written as
\begin{equation}
T_+(x)= e^{i \sqrt{3}\omega a x}z \Phi(\bar{a}+1, 2, z) \nonumber
\end{equation}
with $z=-2i \sqrt{3}\omega a (x+1)$ and $\bar{a}=-\frac{i}{\sqrt{3}}$.
On the other hand, for $T_-$
\begin{equation}
c_2=-1, \qquad b_2=1, \qquad c_0= -3a^2 \omega^2, \qquad b_0=3a^2 \omega^2- 2a \omega. \nonumber
\end{equation}
In that case the solution for $T_-$ can be written as
\begin{equation}
T_-(x)= e^{-i \sqrt{3}\omega a x}\zeta \Phi(\bar{a}+1, 2, \zeta) \nonumber
\end{equation}
with $\zeta=-2i \sqrt{3}\omega a (x-1)$.

\end{document}